# Quantifying the Emergence of Selection Prior to Biological Evolution

Michael Jirasek, Abhishek Sharma, Mary Wong, Jennifer Munro, Leroy Cronin*

School of Chemistry, The University of Glasgow, University Avenue, Glasgow G12 8QQ, UK

*Corresponding author email: Lee.Cronin@glasgow.ac.uk

## Abstract

Selection is central to biological evolution, yet there has been no general experimental framework for quantifying selection in chemical systems before life. Here we demonstrate that selection in a prebiological chemical system can be directly quantified. Assembly Theory predicts that selection corresponds to a transition from undirected to directed exploration of chemical possibility space, measurable through the amount of Assembly, $A$, which integrates molecular assembly index with observed copy number. By analysing peptide ensembles produced under diverse polymerisation conditions, we show that undirected reactions explore sequence space almost uniformly, yielding exploration ratios of 0.85-0.95, whereas reactions influenced by evolved proteases generate markedly lower ratios (0.51-0.75) and elevated $A$, consistent with selective reinforcement of specific assembly pathways. Across multiple environments and amino-acid combinations, the exploration ratio and ensemble assembly $A$ robustly distinguish directed from undirected exploration, establishing a general, experimentally tractable metric for detecting and measuring selection in chemical evolution.

## Introduction

A central open question in the origins of life is how selection first emerges in chemical systems before replication[1], heredity, and Darwinian evolution exist[2,3]. Although many models describe how biological evolution proceeds once genomes arise[4–6], we still lack a universal, experimentally grounded way to detect or quantify selection in purely chemical processes. Without such a measure, the transition from chemistry to biology remains conceptually unresolved and empirically inaccessible. Assembly Theory (AT)[7] provides a route to quantify selection directly from the distributions of objects produced by a chemical system. By defining objects through the minimal number of steps required to construct them, their assembly index, and by integrating this with their observed copy numbers through the assembly equation (1).

$$A = \sum_{i \epsilon n_i > 1}^{N} e^{a_i} \left(\frac{n_i}{N_T}\right) \quad (1)$$

AT predicts a measurable signature of selection: directed processes should generate ensembles enriched in high-assembly objects and constrained subsets of their combinatorial space, whereas undirected processes should explore that space broadly, see Figure 1[8,9]. A second quantity, the exploration ratio, captures how completely a system samples its Joint Assembly Space and therefore reports the degree of directedness.

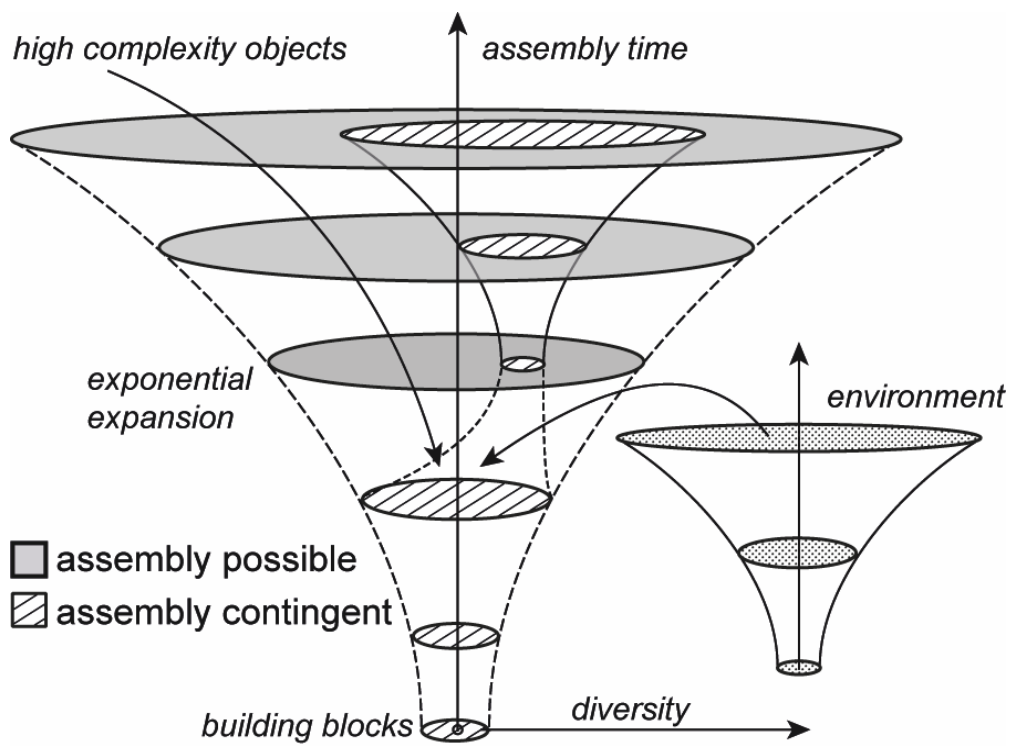

**Figure 1**. Pictographic representation of expanding assembly space with assembly index. Here, we assume that a physical system has a natural propensity to increase in assembly such that assembly index is a monotonic as a function of time. The undirected exploration of the space represents the outer bounds (assembly possible). As an outcome of selection, directness emerges leading the limited exploration of the space (assembly contingent). Here, we represent the emergence of selection in an undirected process as an outcome of interaction with the environment.

A central question is how novel, functional catalysts could arise spontaneously through environmental interactions. Although theoretical studies have examined the timescales and dynamics by which functional molecules might emerge from random chemical "seeds,"[10–13] such processes remain difficult to recognise experimentally without a quantitative measure of selection. Yet understanding when and how function first appears is essential for explaining the origin of life and assessing the likelihood of life elsewhere[14]. Only a few artificial molecular replicators are known, typically relying on aggregation or templating mechanisms[15–17]. Here we show that these predictions, i.e. that AT predicts a measurable signature of selection, hold experimentally. Using peptide formation as a model open-ended chemical system[8], we demonstrate that undirected polymerisation yields exploration ratios approaching unity, consistent with near-uniform traversal of sequence space, while reactions influenced by evolved proteases exhibit sharply reduced ratios and elevated ensemble assembly $A$, revealing selective reinforcement of specific construction pathways. This establishes that a mechanism produced by biological evolution (the protease) can be used as a calibrated environmental constraint, allowing selection to be imposed and quantitatively detected in an otherwise abiotic chemical system. However, the results are general and are not dependent on the enzyme or mechanism of selection. Together, these results show for the first time, that selection in a chemical system can be directly measured from first principles, providing a general, quantitative framework for identifying the onset of directedness in chemical evolution and offering a new experimental route to probe the earliest steps toward biological organisation.

To validate any proposed quantification of selection, it is necessary to test it against a system in which the presence of selection is independently known. Enzymes provide precisely such a reference. Enzymes are molecular artefacts of biological evolution whose sequence-specific constraints are well characterised and quantifiable. By introducing evolved enzymes into an otherwise open, abiotic peptide-forming system, we deliberately embed a known product of biological selection as an environmental constraint. Any valid, general measurement of selection must therefore detect a systematic change in ensemble structure under these conditions, arising specifically from the evolutionary constraints encoded in the enzyme. Failure to do so would falsify the framework.

**Computational Assembly Theory applied to peptide sequences**

In Assembly Theory, the minimal number of steps required to construct an object from defined building blocks, the assembly index, captures the baseline constraint for its formation. For ensembles of objects, the Joint Assembly Space (JAS) represents the minimal set of recursive steps required to construct all observed members. For small molecules this can be computed directly; for larger ensembles, such as peptide mixtures, we approximate the JAS, that is the joint assembly paths of peptides, as the graph union of individual minimal construction paths. For amino acids, assembly indices are derived from their minimal molecular construction pathways. Peptide sequences are then treated modularly by considering amino acids as building blocks connected through amide bonds. This allows the assembly index of a peptide to be approximated as $a_P = a_{JAS} + a_S$ where $a_{JAS}$ is the overall assembly index of the JAS of the amino acids present in the peptide chain, and $a_S$ is the assembly index of the peptide sequence represented by a string, see Figure 2.

Assembly Theory partitions combinatorial space into assembly universe (all abstract possibilities), assembly possible (chemically realisable sequences), assembly contingent (the subset shaped by selection), and assembly observed (the experimentally detected subset). Undirected processes are expected to populate assembly possible broadly and nearly uniformly, whereas directed processes selectively reinforce particular pathways, producing fewer unique sequences with biased distribution across assembly indices. Two metrics emerge from this framework: exploration ratio and ensemble assembly. For a given set of observed objects, the JAS contains observed objects themselves and contingent objects which have not been observed but required for constructing observed objects along the minimal path. The exploration ratio is defined as the ratio of the number of observed objects to the total number of objects which includes both observed and contingent objects. The exploration ratio quantifies how fully the observed ensemble samples its JAS, distinguishing broad, undirected exploration from selective traversal. The ensemble assembly $A$ integrates assembly index and copy number, providing a cumulative measure of the selectivity required to generate the observed products. Together, these quantities provide an experimental route to determine whether a peptide-forming system behaves in an undirected or directed manner.

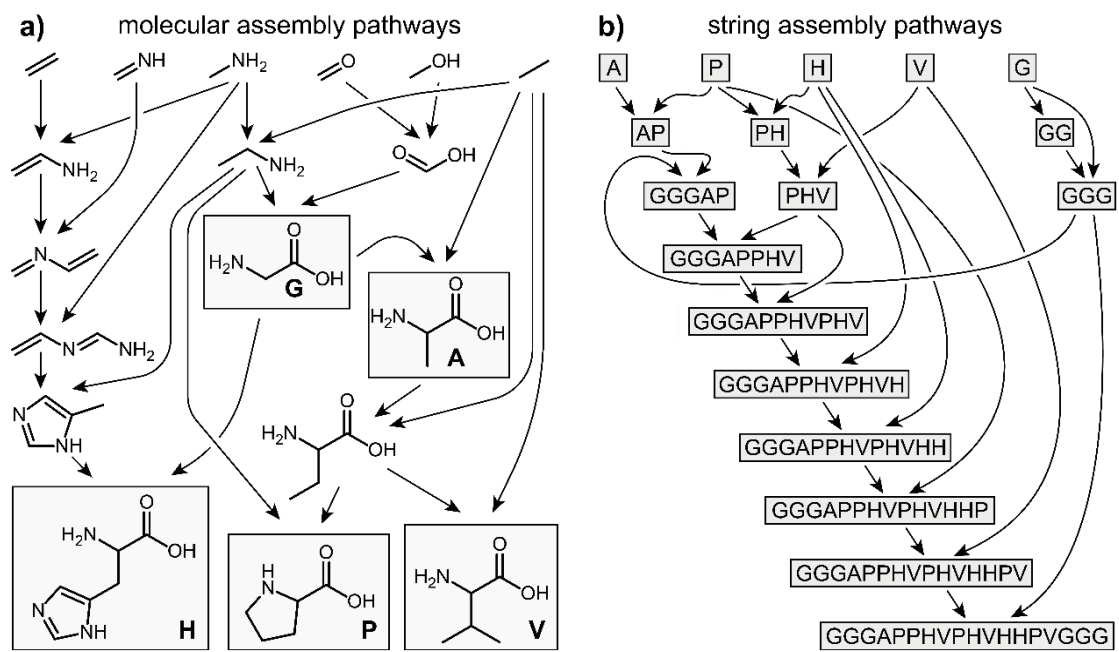

**Figure 2**. a) Joint Assembly Space of five amino acids (glycine, alanine, proline, valine, and histidine) at molecular scale assuming various chemical bonds as a building blocks. b) Assembly pathway of a peptide sequence (GGGAPPHVPHVHHPVGGG) assuming individual characters representing amino acids as building blocks. In AT, a key postulate is that the shortest path to construct an observed object from building blocks. The total assembly index of an unbranched peptide chain can be approximated as $a_P = a_{JAS} + a_S$ where $a_{JAS}$ is the overall assembly index of the JAS of the amino acids present in the peptide chain, and $a_S$ is the assembly index of the peptide sequence represented by a string.

To explore these ideas computationally, we consider five different monomers $\{a, b, c, d, e\}$, and randomly combine them so that once a sequence is formed it can be used recursively. The emerging combinatorial space of sequences therefore represents an undirected exploration without any restriction on copy number. This means that once an object is formed it can be used recursively to create new objects indefinitely, leading to extremely fast exploration of the combinatorial space, see Figure 3a-c. Additionally, without any restriction from copy number, a finite set of stochastic events along the recursive pathways could substantially change the structure of emerging assembly space.

However, combinatorial expansion in the chemical systems is often limited by mass transfer kinetics, and copy number plays a key role. Here, we represent the unexplored assembly space by a stochastic reaction network in which objects are generated recursively through random combinations with and without selective bias as an outcome of environmental interactions. Previously, rule-based chemical space exploration has been used to define chemistry as a recursive expanding reaction network generated by iterative application of fixed reaction rules.[18,19]

So, as a phenomenological model inspired by experimental design, we perform an undirected exploration by starting with five monomers, each with $10^6$ copies, and at each sequence combination step we consider selecting the sequences weighted by their current copy numbers for $10^6$ steps. By limiting combinatorial expansion by copy number, we observed a substantial reduction in the distribution of observed assembly indices, which confirms what we might expect experimentally. Additionally, we estimated the similarity in the observed sequences between multiple runs by estimating the average of intersection-over-union across all pairwise combinations, where the potential space is completely explored among all independent runs at lower assembly indices. The expected deviation in similarity occurs at higher assembly indices where the associated copy number of sequences is very low, see Figure 3d-e.

In the experimental design, cysteine protease introduces environmental selection pressure, biasing the exploration of a vast peptide assembly space towards preferential combinations. Inspired by this, as a phenomenological model, we introduced selection as an environmental effect, where sequences with the end terminal "$b$" are made more favourable for reaction compared to others. This was achieved by introducing a selection factor ($10^2$) which increased the selection probability of the objects with the end terminal "$b$". By running the calculations as in the undirected exploration, we observed sequences with higher assembly indices as well as an increase in similarity at higher assembly indices due to the high usage of "$b$". This suggests that in the case of directed exploration shown here, even if the sequences become more periodic, the system is still capable of increasing the net complexity of the observed ensemble of sequences, see Figure 3f-g.

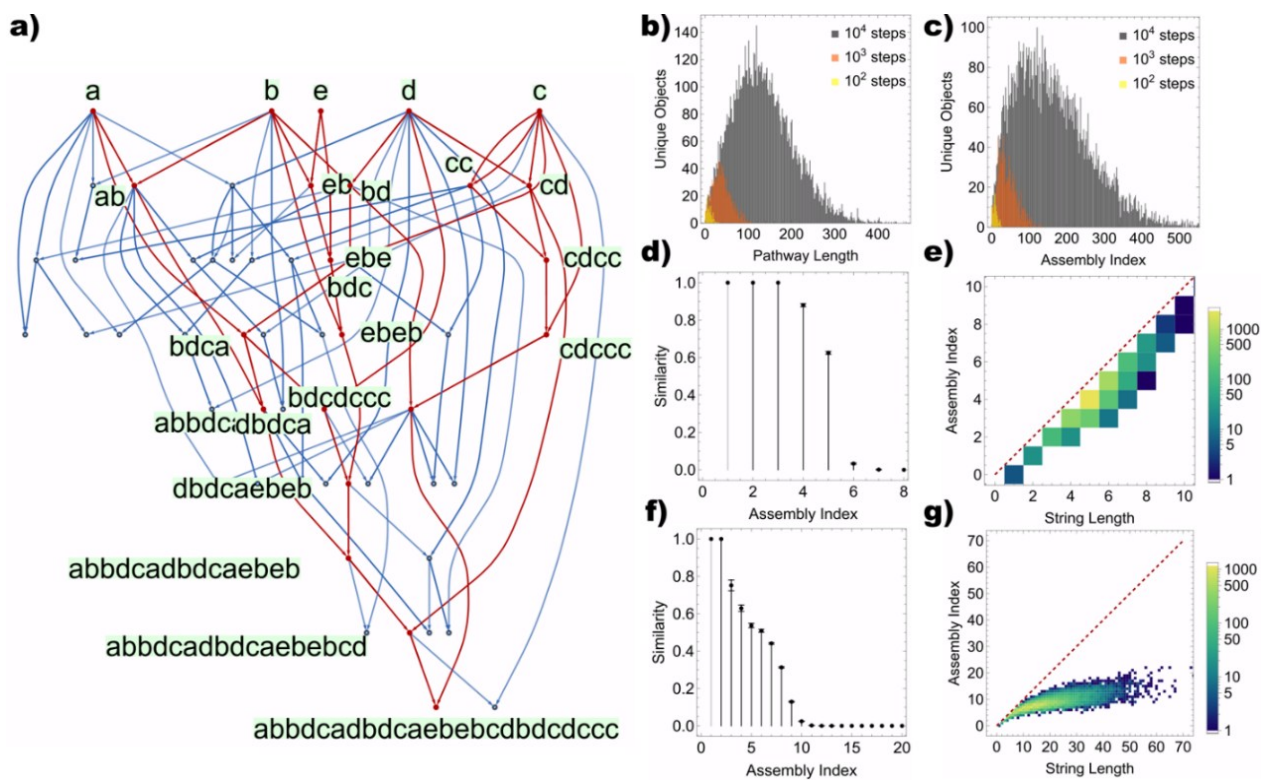

**Figure 3.** Computational experiments on expanding combinatorial space of sequences. a) Expanding space of sequences generated from random combination of sequences with the pool at 50 steps. The pathway of generating the sequence abbdcadbdcaebebcdbdcdccc is highlighted in red. b) and c) shown the pathway length and assembly index of the generated sequences at different timesteps without any limitation on copy number (assembly index calculations with a timeout of 600 seconds). d) shows the similarity between individual runs during an undirected exploration of sequence space weighted by copy number. e) shows the two-dimensional distribution of sequences considering their assembly indices and string lengths for a single run of undirected exploration. f) shows the similarity between individual runs in a directed exploration of sequence space weighted by copy number. The directed exploration was introduced by biasing the weights of selecting sequences with "b" at the endpoints. g) shows the two-dimensional distribution of sequences considering their assembly indices and string lengths for a single run of directed exploration.

## Experimental Assembly Theory Applied to Reactions of Amino Acids

To quantify the degree of selection in experimental systems, we set up wet-lab experiments to explore the combinatorial explosion found in the peptide sequence space formed by the reaction of amino acid monomers, each with different constraints and activation modes. This chemical system allowed us to realise purely undirected exploration, as well as to introduce selection by external factors such as the presence of specific catalysts constructed form the same building blocks. In this regard, enzymatic catalysis was not chosen as an arbitrary source of bias, but as the only currently experimentally accessible case in which selection is known *a priori* to be present, independent of Assembly Theory.

The amount of selection present in the combinatorial explosion can then be quantified using the exploration ratio and assembly $A$. As default, we choose a space of 5 amino acids, glycine (G), L-alanine (A), L-valine (V), L-histidine (H) and L-proline (P) as unprotected amino acids for heat- and coupling agent carbonyldiimidazole (CDI)- activated polymerisation, and the associated methyl esters in the case of enzymes facilitated polymerisation. The amino acids were chosen to be as small as possible (to avoid the formation of oligomers with very large masses), to be soluble in water, and with unique, differentiable exact masses, suitable for mass spectrometry analysis. In several broader substrate screens, we have also used in a larger portfolio of building blocks such as L-phenylalanine (F), L-glutamic acid (E), L-aspartic acid (D), L-cysteine (C), L-serine (S), L-ornithine (O), and L-tryptophan (W).

To realise undirected exploration under different environmental constraints, we polymerised amino acids using three activation modes: heat-driven wet-dry cycling, CDI-mediated coupling in aqueous solution, and CDI-mediated coupling in the solid state. Wet-dry cycling was performed robotically (Opentrons) using XDL[20] procedures, with 50 mM amino acid mixtures heated at 130 °C for 4 h, resuspended, sampled (15%), replenished with fresh monomers, and re-annealed. Solution-phase CDI activation involved mixing amino acids (0.2 M) with CDI (0.5 M) at 40 °C for 1 h, followed by sampling and replenishment. Solid-state CDI coupling was performed in a ball mill ($ZrO_2$ media) at 25 Hz for 90 min, with CDI re-added between cycles. All undirected conditions produced steadily expanding peptide ensembles. To impose directed exploration, we used sequence-selective proteases: papain, bromelain, trypsin, and chymotrypsin, to catalyse peptide formation from amino acid methyl esters in the solid state. Following literature protocols[21,22], amino acid esters were milled with enzyme (10 wt %), L-cysteine for cysteine proteases (5 wt %), and sodium carbonate (1 eq.), with samples taken after each milling cycle.

## Reaction mixture analysis

Samples were resuspended in water, filtered, and analysed by HPLC–MS/MS. Data-dependent acquisition selected the 20 most intense MS1 features for HCD fragmentation at multiple collision energies. MS/MS data were processed using the OLIGOSS software[23], which assigns peptide sequences by matching parent masses to all plausible formulas, validating them through exhaustive b/y fragment searches[24]. Sequences were accepted only when ≥70% of expected fragments were detected, ensuring confident annotation of peptides present at sufficient abundance, see Figure 4.

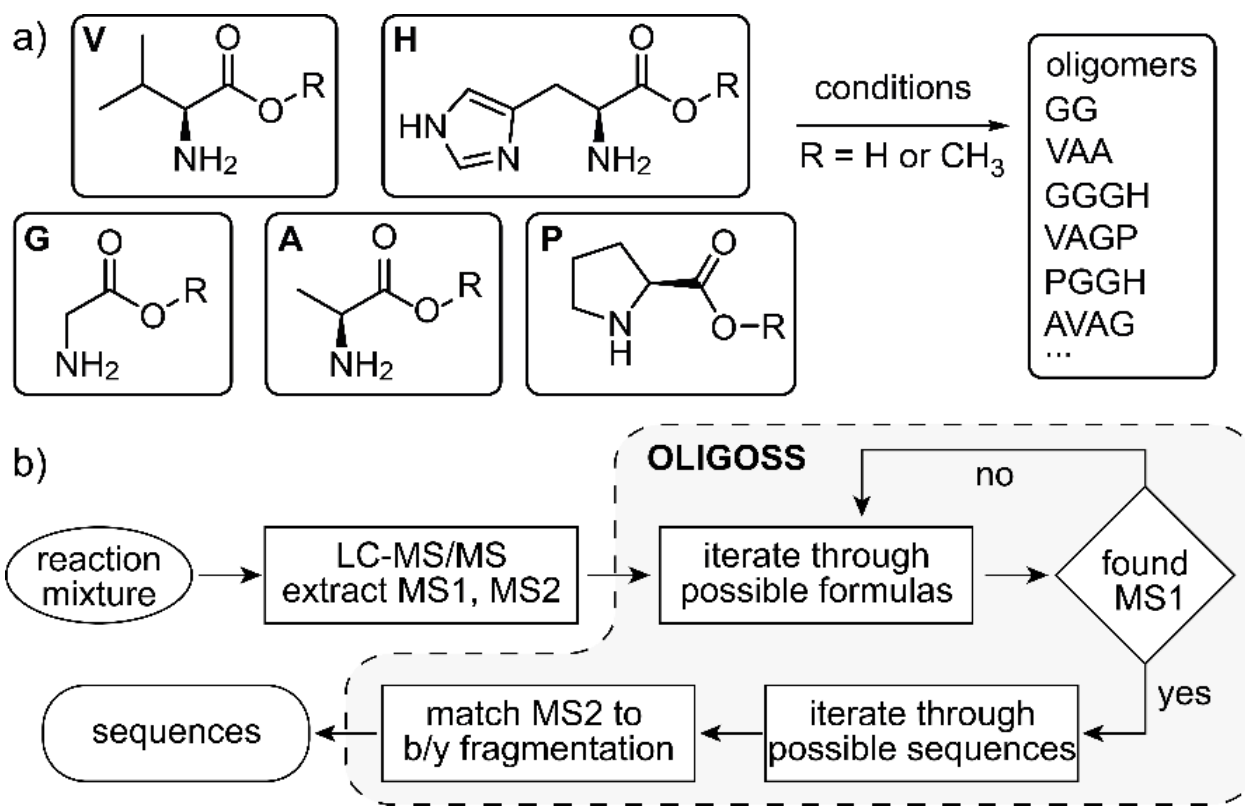

**Figure 4**. Experimental pipeline for experimentally generating and quantifying formation of peptides in an open-ended process. a) Experimental combinatorial explosion is released by providing energy for the reaction, either by heat, coupling agent or under suitable condition catalysed by an enzyme. b) Analysis of a sample using HPLC and MS/MS fragmentation. Found fragments are used, using OLIGOSS,[23] to annotate observed peptides.

To compute the ensemble assembly $A$, we estimated the polymer assembly of each sequence defined as a string with amino acids as building blocks and assigned copy numbers. Because ESI ionisation efficiencies differ widely among peptides and intensities are not quantitatively comparable, we treated all detected species as having equal effective copy number, consistent with their being above the detection threshold[25,26].

For each sequence, the shortest assembly pathway was generated[27], and the assembly path was stored for further analysis. For each ensemble, the JAS path was approximated as the union of the individual paths. Using the generated JAS, the exploration ratio, approximated ensemble complexity, and diversity (measured as the number of unique objects observed) were calculated using Python code, for further information see Code Availability Section below.

## Analysis and Discussion

To evaluate exploration behaviour across recursive cycles, we carried out polymerisation experiments under varied conditions, building-block sets, and activation modes. Exploration ratios and ensemble assembly $A$, calculated using the assembly equation(1), were measured for each run, with all key experiments performed in triplicate and blanks analysed to exclude artefacts. Exploration ratios were computed independently for each condition, with error bars representing standard deviations. Across all undirected polymerisation modes, solution-phase CDI, wet–dry cycling, and solid-state CDI, we observed consistently high exploration ratios (0.85–0.95) and steadily increasing peptide diversity. Solution-phase CDI showed the most extensive undirected exploration (ratios near 1.0), while wet–dry cycling and solid-state CDI both produced ratios around 0.9 across multiple amino-acid combinations, demonstrating that undirected behaviour is robust to reaction environment and composition, see Figure 5. Protease-mediated polymerisation produced a distinct pattern. Cysteine proteases (papain, bromelain) began with low exploration ratios (~0.5) that rose gradually with cycle number, while serine proteases (trypsin, chymotrypsin) maintained intermediate ratios around 0.7 with lower product diversity. These differences reflect enzyme-specific sequence preferences. Ensemble assembly $A$ further separated directed from undirected systems: protease reactions consistently showed steeper increases in $A$ per unique object than CDI-driven reactions. Overall, directed processes produced exploration ratios of 0.55–0.8, while undirected processes remained in the 0.85–0.95 range. This consistent divergence across chemistries and environments demonstrates that the exploration ratio and ensemble assembly *A* reliably distinguish directed from undirected chemical exploration. Crucially, the reduction in exploration ratio and increase in ensemble Assembly are not generic consequences of catalysis, but reveal quantitatively the evolutionary history encoded in the enzyme structure. Assembly therefore detects not merely chemical bias, but the imprint of past selection acting through constraints at the molecular level.

According to peptide database MEROPS,[28] papain exhibits strong and distinctive substrate preferences across its active-site subsites, conventionally numbered P1, P2, … and P1′, P2′, … respectively for amino acid residue positions N-terminal and C-terminal to the cleaved/formed peptide bond. At P2, it favours valine, alanine, and proline, indicating good accommodation of branched aliphatic residues. At P1, glycine is strongly preferred and proline is poorly tolerated, reflecting the need for small, flexible residues at the scissile bond. The P1′ position similarly favours alanine and glycine, with proline completely blocking reactivity. P2′ is more permissive, tolerating glycine and proline well and showing moderate preference for histidine. These patterns explain the selective sequence enrichment observed in papain-mediated polymerisation.

To better understand why the exploration ratio differs and how it reflects the selection involved in the discovery process, we closely examine the JAS using comparable examples from the G, A, V, H, and P amino acid space, and we analyse single observations of peptides with roughly the same number of unique sequences. One dataset represents undirected exploration (using CDI), and the other represents directed exploration (using papain), see Figure 6. In the undirected exploration, the distribution of follows a pattern resembling a normal distribution, rapidly decreasing as MA increases. In contrast, directed exploration produces a distribution skewed toward larger MAs. This is because object-specific rules emerge during directed exploration, where species evolve higher-order specificity. For instance, papain protease exhibits specificity beyond the initial monomers, with the likelihood of a reaction being influenced by interactions up to the fourth amino acid in the enzyme groove.

A more detailed overview of various experiments is summarised in **Table 1.**

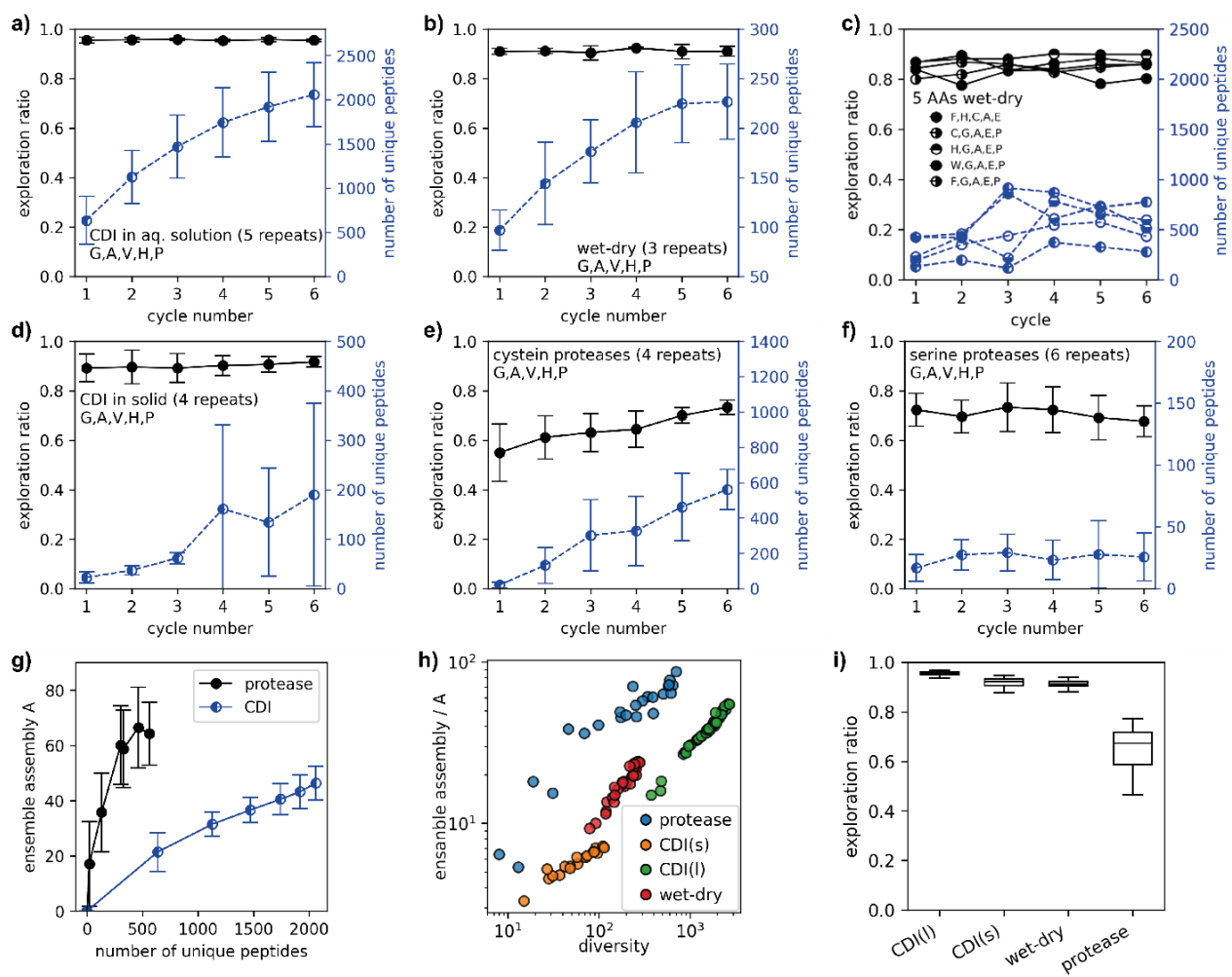

**Figure 5**. Comparison of different metrics of open-ended systems in presence or absence of selection. a-f) Trajectories of exploration ratio over 6 cycles of consecutive polymerisation, presented with error bars representing standard deviation, where high (0.9-1.0) or low (0.55-0.75) exploration ratio demonstrates absence or presence of selection, respectively. a) Exploration ratio (black) and number of unique objects discovered (blue) of polymerisation of G, A, V, H, P building blocks in solution facilitated by CDI; b) facilitated by wet-dry cycling; c) facilitated by wet-dry cycling using various building blocks; d) facilitating by CDI in solid state. e-f) Exploration ratio (black) and number of unique objects discovered (blue) of polymerisation of G, A, V, H, P methyl ester building blocks facilitated by cysteine (e) and serine protease (f) exhibiting lower exploration ratio. g) Assembly $A$ per number of unique objects of ensembles generated with proteases (black) and CDI polymerisation (blue) of G, A, V, H, P showing a clear distinction in slope of $A$/diversity. Overall comparison of ensembles generated by polymerisation facilitated with coupling agent in solid, coupling agent in solution, heat (wet-dry cycles), and proteases, comparing h) $A$ for all ensembles and i) exploration ratios surveyed as a box plot.

To confirm these patterns were not specific to our chosen amino acids, we tested additional compositions. We visualised the JAS of the ensemble, where red nodes represent inferred but unobserved peptides, while white nodes indicate experimentally observed peptides. The JAS of the ensemble produced by directed exploration shows a larger number of contingent, unobserved

– 12 –

peptides, as well as a distribution of observed products shifted toward larger molecular assemblies. Next, we performed bigram transition analysis of observed sequences, visualizing how often each amino acid pair is followed by another amino acid pair within four-residue segments.

| type | Composition / amino acids | diversity | ER(first) | ER(last) | cycles |
|---|---|---|---|---|---|
| CDI (aq.) | G,A,V,H,P | 383–1846 | 0.96 ± 0.00 | 0.95 ± 0.00 | 8 |
| CDI (aq.) | G,A,V,H,P [31:17:18:33:1] | 137–1809 | 0.78 ± 0.01 | 0.80 ± 0.02 | 11 |
| CDI (aq.) | G,A,V,H,P [32:16:15:31:6] | 187–1591 | 0.79 ± 0.04 | 0.77 ± 0.01 | 11 |
| CDI (aq.) | G,A,V,H,P [23:39:4:21:13] | 191–1639 | 0.78 ± 0.02 | 0.78 ± 0.03 | 11 |
| CDI (s.) | G,A,V,H,P | 31–114 | 0.93 ± 0.02 | 0.93 ± 0.01 | 6 |
| wet-dry | H,G,A,E,P | 192–574 | 0.88 ± 0.01 | 0.90 ± 0.00 | 6 |
| wet-dry | F,G,A,E,P | 427–859 | 0.86 ± 0.01 | 0.85 ± 0.01 | 6 |
| wet-dry | F,H,G,A,E | 116–374 | 0.88 ± 0.01 | 0.87 ± 0.01 | 6 |
| wet-dry | C,G,A,E,P | 218–785 | 0.81 ± 0.01 | 0.86 ± 0.00 | 6 |
| wet-dry | W,G,A,E,P | 421–916 | 0.81 ± 0.03 | 0.79 ± 0.01 | 6 |
| wet-dry | V,O,D | 81–149 | 0.93 ± 0.01 | 0.92 ± 0.02 | 10 |
| wet-dry | G,D,H,S | 501–896 | 0.96 ± 0.00 | 0.94 ± 0.01 | 7 |
| protease | Trypsin, G,A,V,H,P | 61–99 | 0.71 ± 0.06 | 0.69 ± 0.05 | 6 |
| protease | Chymotrypsin, G,A,V,H,P | 37–126 | 0.74 ± 0.06 | 0.75 ± 0.03 | 6 |
| protease | Papain, G,V,A,H,P | 16–822 | 0.51 ± 0.01 | 0.73 ± 0.01 | 6 |
| protease | Bromelain, G,V,A,H,P | 26–724 | 0.64 ± 0.08 | 0.69 ± 0.02 | 6 |

**Table 1**. Survey of various representative experiments and averaged values of the diversity and exploration ratios. Diversity is number of unique peptide sequences observed. For full list of all experiments see **SI Section 3**.

Each matrix shows how often specific two-amino-acid combinations are followed by other two-amino-acid combinations within longer peptide sequences, effectively mapping the compositional landscape of four-residue segments generated through distinct experimental approaches. Papain-generated sequences show a highly selective pattern that directly mirrors the enzyme's known substrate selectivity preferences - abundant alanine and valine combinations (AA→VG, or VA→AG), moderate glycine representation, and near-complete absence of proline-containing sequences. This selectivity pattern reflects papain's ability to discriminate between favourable and unfavourable substrates even in reverse synthesis reactions, yielding higher-order selectivity that extends beyond simple amino acid-amino acid preferences to encompass coordinated selectivity across multiple amino acid residues.

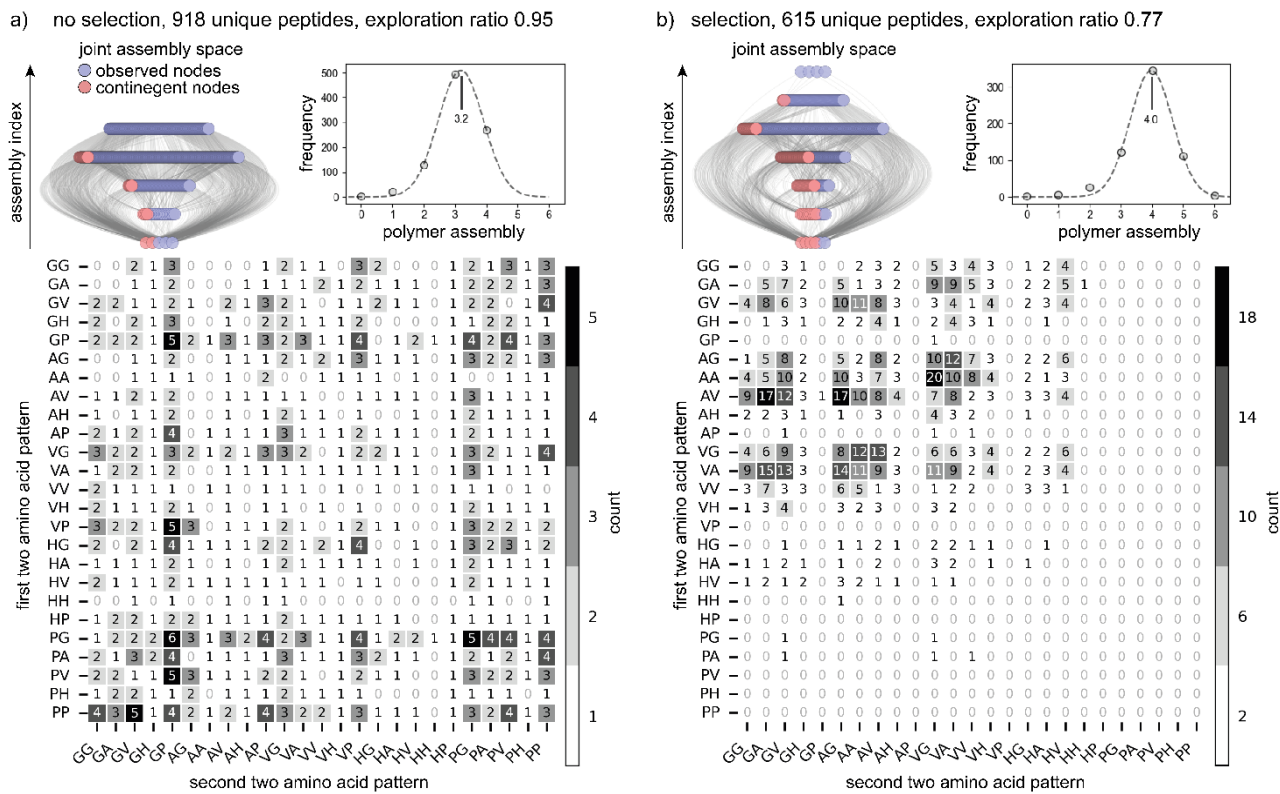

**Figure 6**. Comparison of two datasets produced by an open-ended process in the sequence space of G, A, V, H, and P in the absence of selection (a) (ensemble generated by undirected process (CDI)) and in the presence of selection (b) (less undirected process (enzyme papain) in solid state using ball mill). The joint assembly space of the ensemble shows a larger number of contingent, unobserved nodes (in red), as well as a distribution of observed products shifted toward larger molecular assemblies. Below are bigram matrices of occurrences of specific 4-amino acid sequences. The case of no selection has more homogeneously distributed occurrences compared to the case of selection, where specific higher-order rules dictate more specific distribution.

In contrast, CDI-mediated coupling displays a more uniform distribution across all amino acid combinations, with no strong selectivity bias and notably higher representation of proline-containing sequences. The inverse relationship between these experimental conditions, where papain-generated sequences show high alanine/valine and low proline content, while CDI-generated sequences show more balanced distribution, including significant proline representation which provides evidence that papain's substrate specificity effectively serves as a molecular filter that enriches for sequences containing its preferred amino acid combinations while depleting those containing disfavoured residues like proline.

The transition from undirected to directed exploration, that we have quantified, may represent a critical step in the measurement of the emergence of biological complexity from chemical systems. The exploration ratio provides a quantitative metric for this transition that could be applied to evaluate different prebiotic scenarios. While any chemical condition may bias reaction rates, not all biases constitute selection. In Assembly Theory, selection is defined by the emergence of directed exploration through recursive reinforcement of construction pathways, manifested as constrained traversal of the Joint Assembly Space. The enzymatic systems studied here satisfy this criterion, whereas undirected coupling reactions, despite differing kinetics do not. This is important since this approach could be vital to explore and understand how the environment applies constraints at the molecular level for evolution, e.g. (mineral surfaces, temperature fluctuations, pH gradients) are more conducive to the emergence of selection in primordial chemical systems. Furthermore, our methodology offers a potential approach for detecting biosignatures in samples of unknown origin, as the presence of selection-driven molecular distributions (characterized by low exploration ratios) may indicate biological or pre-biological processes.

## Conclusions

Assembly Theory provides a quantitative framework for detecting the transition from undirected to directed exploration in chemical systems. Here we show that this transition can be measured experimentally in peptide-forming reactions. Analysis of the Joint Assembly Space (JAS) demonstrates that the exploration ratio reliably distinguishes undirected polymerisation, which yields ratios near 0.9, from protease-mediated reactions, which produce markedly lower values (0.62–0.75) consistent with selective reinforcement of specific pathways. Ensemble assembly $A$ further separates these regimes: protease-catalysed reactions exhibit substantially higher $A$ per unique product, revealing enrichment in higher-assembly molecules. Bigram analysis confirms that enzymatic systems impose coordinated multi-residue selectivity, generating characteristic sequence biases that directly reflect substrate preferences. Together, these quantitative and structural signatures show that directed exploration leaves a measurable imprint on the combinatorial landscape of peptide ensembles. Undirected systems produce distributions consistent with combinatorial expectations, whereas directed systems skew strongly toward specific, higher-complexity assemblies. Although more complex chemistries may present additional challenges, the peptide systems studied here demonstrate that selection can be experimentally detected and quantified using the exploration ratio and ensemble assembly. This establishes a general approach for probing the emergence of directedness in chemical evolution, with implications for origins-of-life research, agnostic biosignatures, and the discovery of functional chemical systems through guided exploration of vast molecular spaces.

## Code availability

The code for calculating assembly index and pathways for experimentally observed sequences was previously published,[27] and is available at https://github.com/croningp/assembly_go. Additionally, for sequences generated from simulations, newly developed[29] assembly pathway calculator implementation was used and is available at https://github.com/croningp/assemblycpp-v5. The code developed for processing experimental data for this study is openly available on GitHub at https://github.com/croningp/peptide_selection under MIT licence. The codebase contains generation of JAS, calculations of exploration ratio, approximated ensemble complexity, and diversity (measured as the number of unique objects observed), and generation of figures. The analysis was performed using Python 3.12 and library NetworkX 2.8.8 for graph manipulations. Theoretical simulations of assembly explorations were performed using Wolfram Mathematica 14.

## Data availability

Source data which includes experimental and simulated data used for generating the figures in the main text and supplementary information are available on Zenodo and at https://github.com/croningp/peptide_selection. Additional data such as tandem mass spectra due to large size are available upon request to the corresponding author at Lee.Cronin@glasgow.ac.uk.

## Acknowledgements

We acknowledge financial support from the John Templeton Foundation (grant nos. 61184 and 62231), the Engineering and Physical Sciences Research Council (EPSRC) (grant nos. EP/L023652/1, EP/R01308X/1, EP/S019472/1 and EP/P00153X/1), the Breakthrough Prize Foundation and NASA (Agnostic Biosignatures award no. 80NSSC18K1140), MINECO (project CTQ2017-87392-P), The Eric and Wendy Schmidt Fund For Strategic Innovation (G-21-62939), NIH grant (1UG3TR004136-01) and Alfred P. Sloan foundation (G-2023-21110).

## Author contribution

L.C. conceived the idea and research plan together with M.J. M.J. developed experimental and data processing pipeline with help from AS. M.J., M.W. and J.M. performed the experiments. A.S. and L.C. developed the assembly theory framework and adapted for this work with inputs from M.J. M.J. wrote the manuscript with contributions from all authors. Additionally, we like to thank Dr. Amit Kahana for the feedback on the manuscript and Dr. Alasdair MacLeod for initial experiments.

## References

1. Darwin, C. *On the Origin of Species*. (Macmillan Collector's Library, London, 2017).
2. Nowak, M. A. *Evolutionary Dynamics: Exploring the Equations of Life*. (Belknap Press of Harvard University Press, Cambridge, Mass, 2006).
3. Gánti, T. *The Principles of Life*. (Oxford University Press, Oxford, 2006).

4. Von Kiedrowski, G. A Self-Replicating Hexadeoxynucleotide. *Angew. Chem. Int. Ed. Engl.* **25**, 932–935 (1986).

5. Eigen, M. Selforganization of matter and the evolution of biological macromolecules. *Naturwissenschaften* **58**, 465–523 (1971).

6. Eigen, M., McCaskill, J. & Schuster, P. Molecular quasi-species. *J. Phys. Chem.* **92**, 6881–6891 (1988).

7. Sharma, A. *et al.* Assembly theory explains and quantifies selection and evolution. *Nature* https://doi.org/10.1038/s41586-023-06600-9 (2023) doi:10.1038/s41586-023-06600-9.

8. Taylor, T. Requirements for Open-Ended Evolution in Natural and Artificial Systems. Preprint at https://doi.org/10.48550/ARXIV.1507.07403 (2015).

9. Hochberg, M. E., Marquet, P. A., Boyd, R. & Wagner, A. Innovation: an emerging focus from cells to societies. *Phil. Trans. R. Soc. B* **372**, 20160414 (2017).

10. Joyce, G. F. Directed Evolution of Nucleic Acid Enzymes. *Annu. Rev. Biochem.* **73**, 791–836 (2004).

11. Seelig, B. & Szostak, J. W. Selection and evolution of enzymes from a partially randomized non-catalytic scaffold. *Nature* **448**, 828–831 (2007).

12. Vaidya, N. *et al.* Spontaneous network formation among cooperative RNA replicators. *Nature* **491**, 72–77 (2012).

13. Ameta, S. *et al.* Darwinian properties and their trade-offs in autocatalytic RNA reaction networks. *Nat Commun* **12**, 842 (2021).

14. Patel, B. H., Percivalle, C., Ritson, D. J., Duffy, C. D. & Sutherland, J. D. Common origins of RNA, protein and lipid precursors in a cyanosulfidic protometabolism. *Nature Chem* **7**, 301–307 (2015).

15. Adamski, P. *et al.* From self-replication to replicator systems en route to de novo life. *Nat Rev Chem* **4**, 386–403 (2020).

16. Issac, R. & Chmielewski, J. Approaching Exponential Growth with a Self-Replicating Peptide. *J. Am. Chem. Soc.* **124**, 6808–6809 (2002).

17. Mizuuchi, R., Furubayashi, T. & Ichihashi, N. Evolutionary transition from a single RNA replicator to a multiple replicator network. *Nat Commun* **13**, 1460 (2022).

18. Andersen, J. L., Flamm, C., Merkle, D. & Stadler, P. F. Inferring chemical reaction patterns using rule composition in graph grammars. *J Syst Chem* **4**, 4 (2013).

19. Andersen, J. L., Flamm, C., Merkle, D. & Stadler, P. F. A Software Package for Chemically Inspired Graph Transformation. in *Graph Transformation* (eds Echahed, R. & Minas, M.) vol. 9761 73–88 (Springer International Publishing, Cham, 2016).

20. Rauschen, R., Guy, M., Hein, J. E. & Cronin, L. Universal chemical programming language for robotic synthesis repeatability. *Nat. Synth* **3**, 488–496 (2024).

21. Terada, K. *et al.* Papain-Catalyzed, Sequence-Dependent Polymerization Yields Polypeptides Containing Periodic Histidine Residues. *Macromolecules* **55**, 6992–7002 (2022).

22. Ardila-Fierro, K. J. *et al.* Papain-catalysed mechanochemical synthesis of oligopeptides by milling and twin-screw extrusion: application in the Juliá–Colonna enantioselective epoxidation. *Green Chem.* **20**, 1262–1269 (2018).

23. Doran, D. *et al.* Exploring the sequence space of unknown oligomers and polymers. *Cell Reports Physical Science* **2**, 100685 (2021).

24. Chu, I. K. *et al.* Proposed nomenclature for peptide ion fragmentation. *International Journal of Mass Spectrometry* **390**, 24–27 (2015).

25. Silva, J. C., Gorenstein, M. V., Li, G.-Z., Vissers, J. P. C. & Geromanos, S. J. Absolute Quantification of Proteins by LCMSE. *Molecular & Cellular Proteomics* **5**, 144–156 (2006).

26. Ong, S.-E. & Mann, M. Mass spectrometry–based proteomics turns quantitative. *Nat Chem Biol* **1**, 252–262 (2005).

27. Jirasek, M. *et al.* Investigating and Quantifying Molecular Complexity Using Assembly Theory and Spectroscopy. *ACS Cent. Sci.* **10**, 1054–1064 (2024).

28. Rawlings, N. D. *et al.* The MEROPS database of proteolytic enzymes, their substrates and inhibitors in 2017 and a comparison with peptidases in the PANTHER database. *Nucleic Acids Research* **46**, D624–D632 (2018).

29. Seet, I., Patarroyo, K. Y., Siebert, G., Walker, S. I. & Cronin, L. Rapid Exploration of the Assembly Chemical Space of Molecular Graphs. *J. Chem. Inf. Model.* acs.jcim.5c01964 (2025) doi:10.1021/acs.jcim.5c01964.